\documentclass{sig-alternate}
\usepackage{cite}
\usepackage[usenames,dvipsnames,svgnames,table]{xcolor}
\usepackage{graphicx}
\usepackage{caption}
\usepackage{pbox}
\usepackage{url}
\usepackage{multirow}

\usepackage{tabularx}
\usepackage{multirow}

\usepackage{todonotes}

\usepackage[strict]{changepage}

\usepackage{framed}

\definecolor{formalshade}{rgb}{0.95,0.95,1}

\makeatletter
\newcommand{\interviewquote}[2]{\vspace{-0.2	cm}
 \def\FrameCommand{%
    \hspace{0pt}%
    {\color{MidnightBlue}\vrule width 1.5pt}%
    {\color{white}\vrule width 4pt}%
    \colorbox{white}
  }%
  \MakeFramed{\advance\hsize-\width\FrameRestore}%
  \noindent\hspace{-4.55pt}% disable indenting first paragraph
  \begin{adjustwidth}{}{7pt}
  \footnotesize{"\emph{#1}" -{#2}}\vspace{0.5pt}\end{adjustwidth}\endMakeFramed%
}
\makeatother

\newcommand{\subheading}[1]{\textbf{\emph{#1}}. }

\newcommand{\observation}[1]{
6.56
}

\newcommand{\simpleobservation}[2]{
}

\usepackage{ifthen}

\newboolean{showcomments}
\setboolean{showcomments}{false}  %% toggle
\ifthenelse{\boolean{showcomments}}
  {\newcommand{\note}[2]{
    \fbox{\bfseries\sffamily\scriptsize#1}
    {\sf\small\textit{#2}}
  }
 }
  {\newcommand{\note}[2]{}
   
  }

  \usepackage{tcolorbox}
  \newcommand{\summary}[1]{
  	\vspace{0.25cm}
  	\begin{tcolorbox}
  		\small{
  		#1}
  	\end{tcolorbox}
  }

\begin{document}
%PLv1: \title{The making of the cloud - a 360-degree exploratory study of software development for and in the cloud}
%TF: \title{Developing for the Cloud}
%PLv2:
%\title{The Making of Cloud Applications -- An Empirical Study of How Cloud Software is Built}
\title{The Making of Cloud Applications -- An Empirical Study on Software Development for the Cloud}

%\author{\IEEEauthorblockN{J\"urgen Cito, Philipp Leitner, Thomas Fritz, Harald C. Gall\\
%s.e.a.l. -- software evolution \& architecture lab\\
%	Department of Informatics\\
%              University of Zurich\\
%              Zurich, Switzerland \\
%             \{lastname\}@ifi.uzh.ch}}

\numberofauthors{1} %  in this sample file, there are a *total*
\author{
	% 1st. author
	J\"urgen Cito, Philipp Leitner, Thomas Fritz, Harald Gall\\
%		\affaddr{Department of Informatics}\\
	\affaddr{University of Zurich, Switzerland}\\
	\email{\{lastname\}@ifi.uzh.ch}
}

%\numberofauthors{1} 
%\author{\alignauthor Bernard Rous, Sally Smith, Jack Sheridan, and Bill Jones \\
%	\affaddr{Dept. of Computer Science, University of Greendell} \\ \affaddr{ City, State, Country} \\
%	\email{br@cs.cxcc.edu, ss@cs.cxcx.edu, js@cs.cxcs.edu, bj@ccxc.com}
%} 

\maketitle

\begin{abstract}
Cloud computing is gaining more and more traction as a deployment and provisioning model for software. While a large body of research already covers how to optimally operate a cloud system, we still lack insights into how professional software engineers actually use clouds, and how the cloud impacts development practices. This paper reports on the first systematic study on how software developers build applications in the cloud. We conducted a mixed-method study, consisting of qualitative interviews of 25 professional developers and a quantitative survey with 294 responses. Our results show that adopting the cloud has a profound impact throughout the software development process, as well as on how developers utilize tools and data in their daily work. Among other things, we found that (1) developers need better means to anticipate runtime problems and rigorously define metrics for improved fault localization and (2) the cloud offers an abundance of operational data, however, developers still often rely on their experience and intuition rather than utilizing metrics. From our findings, we extracted a set of guidelines for cloud development and identified challenges for researchers and tool vendors.

\end{abstract}



% \fi
%\IEEEpeerreviewmaketitle

\section{Introduction}
Since its emergence, the cloud has been a rapidly growing area of interest~\cite{buyya:09,Armbrust:2010qy}. Several cloud platforms, such as Amazon's EC2, Microsoft Azure, Google's App Engine, or IBM's Bluemix, are already gaining mainstream adoption. Developing applications on top of cloud services is becoming common practice. Due to the cloud's flexible provisioning of resources, and the ease of offering services online for anyone, the cloud also influences software development practices. For instance, cloud development is often associated with the concept of ``DevOps'', which promotes the convergence of the development and operation of applications~\cite{huettermann:12}.

There is currently significant research interest in how to efficiently manage cloud infrastructures, for instance in terms of energy efficiency~\cite{beloglazov:12} or maximized server utilization~\cite{marshall:11}. Another core area of interest in cloud computing research is its use for high-performance computing in lieu of an expensive computer grid~\cite{iosup:11}. However, so far, there is little systematic research on the consumer side of cloud computing, i.e., how software developers actually develop applications in and for the cloud. Only recently, Barker et al. voiced this concern in a position paper, stating that the academic community ought to conduct more ``user-driven research''~\cite{barker:14}.

In this vein, this paper presents a systematic study on how professional software engineers develop applications on top of cloud infrastructures or platforms. We deliberately cover a broad scope, and analyze how applications are designed, built and deployed,
as well as what technical tools are used for cloud development.
We conducted a mixed-method study consisting of an initial interview study with 16 professional cloud developers, a quantitative survey with 294 respondents, and a second round of interviews with 9 additional professionals to dive deeper into some questions
raised by the survey.
All interview participants work at international companies of widely varying size (from small start-ups to large enterprises), and have diverse backgrounds with professional experience ranging from 3 to 23 years.

In particular, we addressed the following two research questions:

\vspace{0.25cm}

\emph{\textbf{RQ 1}: How does the development and operation of applications
change in a cloud environment?}\\
In the cloud, servers are \emph{volatile}. They are regularly terminated
and re-created, often without direct influence of the cloud developer.
Our study has shown that the concept of API-driven infrastructure-at-scale and this cloud
instance volatility have ripple effects throughout the entire
application development and operations process. They restrict the design of
cloud applications and force developers to heavily rely on  infrastructure
automation, log management, and metrics centralization. While these concepts are
also useful in non-cloud environments, they are mandatory for  successful
application development and operation in the cloud.

\vspace{0.2cm}
	
\emph{\textbf{RQ 2}: What kind of tools and data do developers utilize for
building cloud software?}\\
Based on our research, more data, and more types of data, are utilized in the
cloud, for instance business metrics (e.g., conversion rates) in
addition to system-level data (e.g., CPU utilization).
However, developers struggle to directly interpret and make use of
this additional data, as current metrics are often not actionable for them.
Similarly, cloud developers are in the abstract aware that their design and
implementation decisions have monetary consequences, but in their daily
work, they do not currently think much about the costs of operating their
application in the cloud.

\vspace{0.25cm}

Our research has important implications for cloud developers, researchers, and
vendors of cloud-related tooling. Primarily, due to the volatility of cloud
instances, developers need to accustom themselves to not being able to
directly touch the running application any longer. That is, quick fixes
of production configuration are equally impossible as logging into a server
for debugging. As a research community,
we need to investigate how to best support developers in this task, as well
as analyze how code artefacts related to cloud instance management evolve.
Finally, we have seen that more types of metrics get more and more
important, but they are still not directly actionable for developers. Hence,
we need to research better tooling that brings this data into the daily workflow
of cloud developers.

% 
% Our results show that the cloud has many advantages for software development, for example allowing developers to release features faster. 
% Further, we observe the trend to a more DevOps style of communication, and increased usage of runtime data (most importantly performance data) from the cloud. While we have seen that deploying on a cloud restricts software developers, both technologically and in terms of architectural decisions, these restrictions are often not seen as negative.

% However, given that cloud deployment generally implies a loss of control, cloud developers also need to monitor their application, as well as the cloud, closely. This need for more monitoring goes hand in hand with a tighter collaboration between developers and operators, even though the vision of DevOps is not yet fully realized in most companies. 

%The responses from the 294 survey participants largely confirm our initial observations, showing that development projects in the cloud are more agile and deliver features faster. Performance monitoring and collaboration between development and operations teams are becoming more important. However, we also observe that cloud developers need to deal with restrictions, both technologically and in terms of application design. 

The remainder of the paper is structured as follows. First, we provide some background on cloud computing terminology (Section~\ref{sec:bg}), followed by a discussion of related work in Section~\ref{sec:rw}. We then present the study design (Section~\ref{sec:methodology}), followed by an in-depth summary of our findings (Section~\ref{sec:findings}) and a discussion of the implications resulting from those findings (Section~\ref{sec:implications}). We detail the major threats to validity of our research in Section~\ref{sec:threats}, and conclude the paper in Section~\ref{sec:conc}.
\section{Background}
\label{sec:bg}

\begin{figure}[h!] 
	\centering
	\includegraphics[width=0.4\textwidth]{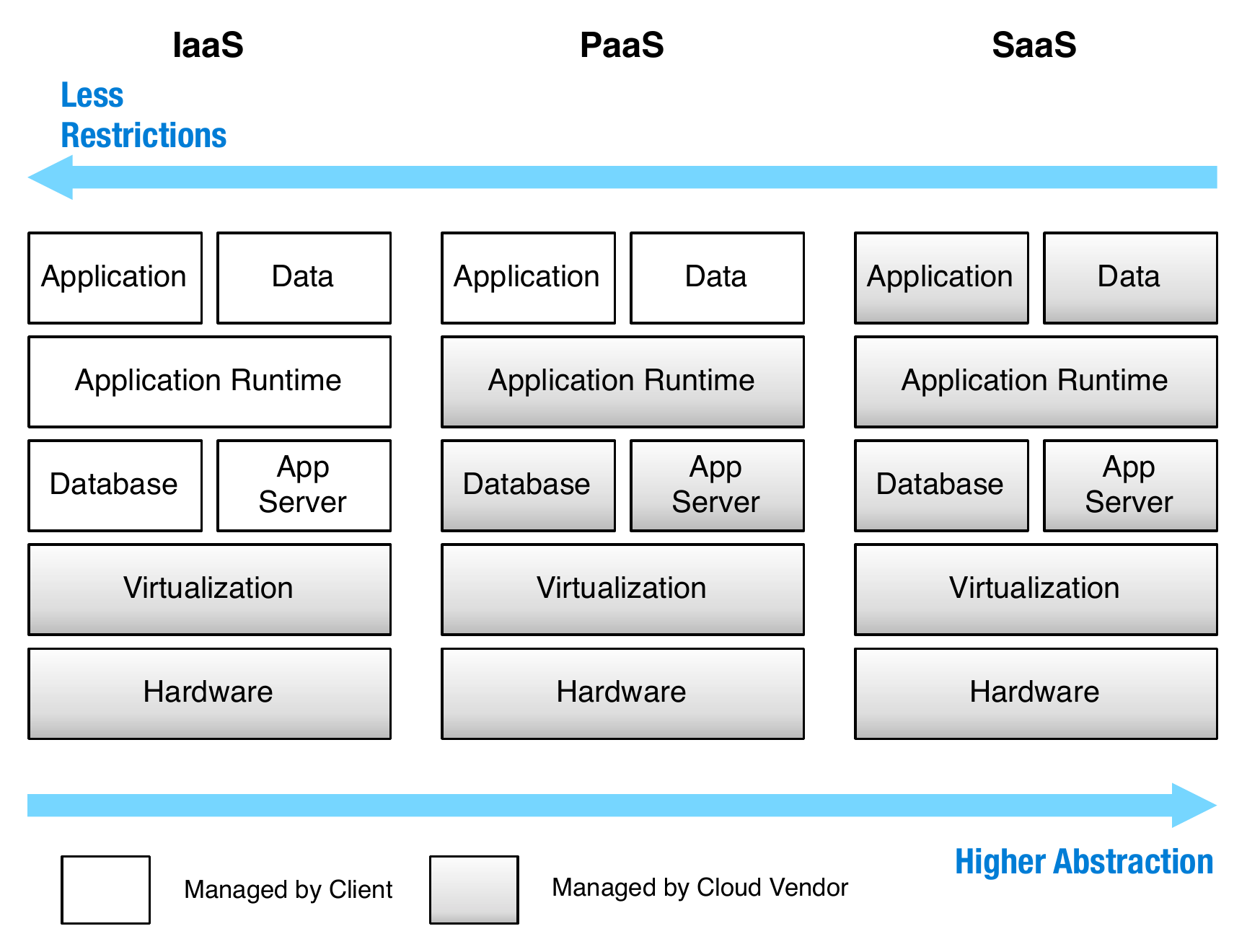}
	\caption{Basic models of cloud computing (following~\cite{mell:13})} 
	\label{fig:def}
\end{figure}

While the term ``cloud computing'' is commonly ill-defined\footnote{Oracle's CEO Larry Ellison once noted jokingly that he cannot think of a single thing Oracle does that is not ``cloud''.}, the research community has widely gravitated towards the NIST definition~\cite{mell:13}. As illustrated in Figure~\ref{fig:def}, this definition considers three levels, each defined by the responsibilities of IT operations provided by the cloud vendor. In an \emph{Infrastructure-as-a-Service (IaaS)} cloud, resources (e.g, computing power, storage, networking) are acquired and released dynamically, typically in the form of virtual machines. IaaS customers do not need to operate physical servers, but are still required to administer their virtual servers, and manage installed software.  \emph{Platform-as-a-Service (PaaS)} clouds represent a higher level of abstraction, and provide entire application runtimes as a service. The PaaS provider manages the hosting environment, and customers only submit their application bundles. They typically do not have access to the physical or virtual servers on which the applications are running. %PaaS customers only communicate with the platform via its API and they are largely relieved from server administration and operation duties. 
Finally, in \emph{Software-as-a-Service (SaaS)}, complete applications are provided as cloud services to end customers. The entire stack, including the application, is handled by the provider. The client is only a user of the service.

PaaS clouds are particularly interesting for software engineers, as they allow  them to solely focus on developing software applications. They typically relieve the developer from having to care about any operations tasks, and handle varying system load transparently via auto-scaling. This ability to adapt to workload changes is referred to as \emph{elasticity}. However, in order to do so, these platforms impose severe restrictions. For instance, they typically only support rather narrowly defined application models (e.g., three-tier Web applications), and require the developer to program against provided APIs. This often also leads to vendor lock-in~\cite{lawton:08}. 

%elasticity is defined as the degree to which a system is able to adapt to workload changes

% Orthogonally to these three service models, the NIST definition of cloud computing also distinguishes between four deployment models. \emph{Private clouds} are cloud systems that are provided for internal use (e.g., inside a larger company), while \emph{public clouds} are open to anybody (although typically not for free). \emph{Hybrid cloud} solutions typically represent a combination of private and public clouds, often in the form of \emph{cloud bursting}~\cite{guo:12} (operating one or more applications primarily from a private cloud, but covering load spikes with resources from a public cloud). Finally, \emph{community clouds} are clouds that are accessible only by a restricted set of users, which often have to ``buy in'' into the system by providing their own resources to other users as well. 
With IaaS, the  idea of Infrastructure-as-Code (IaC) has also started to gain momentum. IaC allows users to define and provision operation environments in version-controlled source code. Essentially, in an IaC project, the entire runtime environment of the application (e.g., IaaS resources, required software packages, configuration) is defined using scripts, which can then be executed by tools such as Opscode Chef\footnote{\url{https://www.getchef.com}}. These scripts allow entire test, staging or production environments to be started without manual interaction. The move towards IaC with its reproducible provisioning has become necessary since cloud applications often consist of a large number of machines that have to be configured automatically to scale horizontally.

Another concept commonly associated with cloud development is DevOps~\cite{huettermann:12}. DevOps describes the convergence of the previously mostly separated tasks of developing an application, and its deployment and operation. In DevOps, software development and operation activities are often handled by the same team, or even by the same engineer. By aligning the goals of development and operations, DevOps aims at improving agility and cooperation.

\section{Related Work}
\label{sec:rw}

There has been a multitude of empirical research on the development of general software applications. For instance, Singer et al. have recently researched how developers use Twitter~\cite{singer:14}. Murphy-Hill et al. have looked at how bugs are fixed~\cite{murphy:13}. However, so far, very little empirical research has been conducted
in the cloud computing domain, even though there are several calls for more research on software development for the cloud.  Barker et al.~\cite{barker:14} recently named ``user-driven research'' as one of the major opportunities for high-impact cloud research. Khajeh-Hosseini et al.~\cite{khajeh:11} stated that the organizational and process-oriented changes implied by adopting the cloud is currently not sufficiently researched. While Mei et al. did not consider software engineering a major challenge for cloud computing in 2008~\cite{mei:08}, they later on provided a whole list of software engineering issues to be tackled by research~\cite{mei:13}.

So far, research in cloud computing has mainly focused on provider-side issues (e.g., relating to server management~\cite{beloglazov:12,marshall:11} or performance measurement\cite{iosup:11}). 
On the client side, some research has been conducted on concrete programming models. A large part of this research deals with data analysis, typically using the Map/Reduce paradigm (e.g.,~\cite{palanisamy:11}). While interesting, these works do not cover the professional software development environment that we address with our study.
Research on cloud programming models for non-scientific contexts is more limited. One example is the jCloudScale framework proposed in~\cite{leitner:12}. jCloudScale is a Java-based middleware that aims to simplify the development of IaaS applications. A similar goal also motivated the research presented in~\cite{jayaram:13}, which investigated an extension of Java RMI for simplifying the development of elastic, cloud-based applications.
%Finally, a .NET-based cloud programming model has been proposed by researchers from Microsoft Research~\cite{bykov:11}. 

One aspect that is already reasonably well-understood in literature is how and when companies choose to adopt cloud computing, and for which reasons. A large-scale survey on this topic has been presented in~\cite{narasimhan:11}. The authors conclude that improved business agility is a larger factor for companies to adopt the cloud than reduced costs. In a second study on cloud adoption in small and medium-sized enterprises~\cite{gupta:13}, the authors conclude that ease of use and convenience is a more important reason for adoption than both, reduced costs and improved business agility. However, both of these studies are concerned primarily with SaaS adoption. That is, they target cloud adoption by end users more than by professional software developers. This is not the case in a related industry study, dubbed the ``DevOps Report''~\cite{devops:14}. This survey garnered over 9200 respondents, praising the DevOps idea as a key enabler of profitable and agile companies. Given that the source of this report is also a major player in the DevOps business, independent scientific evaluation to support these results would be valuable.

None of the work discussed so far has empirically evaluated how cloud software is actually developed in practice. The only work we are aware of that goes into this direction is a (not peer-reviewed) white paper on enterprise software development in the cloud~\cite{shiver:14}. This report is based on a survey with 408 respondents. The report concludes that enterprise developers are largely not yet adopting the cloud, but if they do, they are able to improve time-to-market.
%This is consistent with our results. We have also observed that features can be released to customers much more quickly in the cloud. Further, about 2 out of 3 respondents of our validation survey came from developers employed by companies with 20 employees or less, suggesting the claim that cloud computing is currently of more interest to smaller companies. 
\section{Research Method}
\label{sec:methodology}

%\textbf{THOMAS: Is there a way to put P9$_{PaaS}$ or so into all quotes please!!!}
%
%Study Type / Participants / Company size / platform
%# / IDs // small / enterprise  // PaaS / IaaS / both
%
\newcolumntype{g}{>{\columncolor[gray]{0.9}}l}
\begin{table*}[t] \centering
	\caption{Method and Participants}
	\label{tab:studies}
	\begin{tabular}{l c c c c c c c c c c }
		\hline
		\multirow{2}{*}{\textbf{Study Type}} & \multirow{2}{*}{\textbf{\# Questions}}  & \multicolumn{2}{c}{\textbf{Participants}}  & \multicolumn{2}{c}{\textbf{Platform}} & \multicolumn{2}{c}{\textbf{Company}} & \textbf{Experience} \\
		&  & \# & IDs & PaaS & IaaS & enterprise & small & Avg ($\pm$ StdDev) \\
		\hline
		\rowcolor[HTML]{EFEFEF} 
		\textbf{Interviews} & \textbf{50} & \textbf{25} &  & \textbf{15} & \textbf{10} &  \textbf{14} & \textbf{11} & \textbf{9 years ($\pm$ 6.5)} \\
		\hspace{0.6em} Interview$_1$ & 23 & 16 & P1 - P16 & 13 & 3 &  12 & 4 & 9 years ($\pm$ 7) \\
		\hspace{0.6em} Interview$_2$ & 27 & 9 & D1 - D9 & 3 & 6 & 2 & 7 & 8 years ($\pm$ 6) \\
		\rowcolor[HTML]{EFEFEF} 
		\textbf{Survey} & \textbf{23} & \textbf{294} & & \textbf{103} & \textbf{191} & \textbf{102} & \textbf{192} & \textbf{9 years ($\pm$ 5)} \\
		
		\hline
	\end{tabular}
\end{table*}

To investigate how the cloud influences software development practices, we conducted a study based on techniques found in Grounded Theory~\cite{hoda:12}. Following the recommendations in~\cite{bratthall:02}, we used a mixed methodology consisting of three steps of data collection and iterative phases of data analysis. First, we defined a set of open-ended questions from our research questions and conducted \emph{qualitative interviews} with 16 participants. Second, to further substantiate our findings, we ran a \emph{quantitative survey} and gathered responses from 294 professional software developers. Using open coding, we identified 4 topics of high interest. To gain a better understanding and more details on these topics, we then conducted a second round of \emph{qualitative deep-dive interviews} with 9 professional developers. All interview materials and survey questions can be found on our web site\footnote{http://www.ifi.uzh.ch/seal/people/cito/cloud-developer-study.html}.

\paragraph{Qualitative Interview Study (Interview$_1$)}
\textbf{Protocol.} We conducted semi-structured interviews with software developers that previously already deployed software in the cloud in a professional context. For these interviews, we defined a set of 23 questions based on our research questions. In the interviews, we covered all questions with each participant, but the concrete order of questions followed the ``flow'' of the interview. Interviews were conducted by the first author, either face-to-face on-site of the interviewee or via Skype. Interviews lasted between 30 and 60 minutes, were conducted in either German or English, and were audio-recorded.

\textbf{Participants.}
Interview participants were recruited from industry partners and our personal network. To cover a broad range of cloud development experiences, we recruited participants from both, smaller companies (1 -- 100 employees) and larger enterprises ($>$ 100 employees). Participants had to either deploy on IaaS or PaaS as well as in public and private clouds. Furthermore, we also made sure to recruit some participants that delivered SaaS applications. Overall, we recruited 16 participants (P1 to P16), all male software developers with professional development experience between 3 and 23 years (average of 9 years $\pm$ 7 standard deviation), and from 4 different countries and two continents. Our 16 participants came from 5 different companies---12 participants worked in larger enterprises, 4 in smaller companies.
%---with 13 participants deploying on PaaS and 3 participants deploying on IaaS.
%Participants P6, P7 and P8 are not part of the interview study, as those were used in the exploratory phase of our research.

\textbf{Analysis.} After the interviews, we transcribed all audio recordings. The first two authors then used an open coding technique to iteratively code and categorize participants' statements, resulting in a set of findings. All findings are supported by statements of multiple interview participants. %Due to the inherent heterogeneity of our interviewees, some of these observations relate mostly to developers working in either smaller or larger companies, or with specific cloud models (e.g., IaaS or PaaS).
% since there is a difference between developers deploying on IaaS or PaaS technologies. 

% \subsubsection{Protocol}
% 
% We started off by asking an open-ended and deliberately ambiguous question: 'How is developing applications in the cloud different from developing in a traditional environment?'. As expected, this question sparked a discussion on what the 'cloud' actually is, and what 'traditional' means.
% 
% Afterwards, we addressed each of the topics we derived in the previous phase (see Section \ref{sec:relatedworkstudy}). We followed up on aspects we seemed were interesting and would probably yield new insights. 
% Finally, we thanked the developer and expressed our gratitude with Swiss chocolate. We informed them of our future work with the data and asked for permission to contact them again for clarification that might be needed during the coding and analysis phase. The interviews were voice recorded and then transcribed.
% 
% \subsection{Coding \& Analysis}
% In the third phase, we employed open coding, iterating through our interview transcripts first extracting categories from statements. In further iterations we extracted codes and formulated hypothesis and findings that we report in Section \ref{sec:findings}.

\paragraph{Quantitative Study (Survey)}
\textbf{Protocol.} In the second step of our study, we designed a survey with 32 questions, most of which map to our initial findings. The questions were primarily formulated as statements, asking participants to state their agreement on a five point Likert scale (examples of these questions can be seen in Figure \ref{fig:findings}). To target developers with experience in cloud technologies, we gathered email addresses of GitHub\footnote{http://www.github.com} users that ``follow'' a repository of a number of popular cloud platforms, including Amazon Web Services, Heroku, Google Cloud Platform, CloudFoundry, and OpenStack. We then discarded all users without a public email address in their profile, and contacted the remaining users with a description and link to our online survey. To motivate developers to participate in the survey, we held a raffle for all participants to win one of two 50 USD Amazon gift vouchers. The survey was in English, and took an average of 12.2 minutes to complete.

\textbf{Participants.} We emailed the survey invitation to 2000 GitHub users and gathered a total of 294 responses (response rate of 14.7\%). Of all 294 participants, 192 were employed in smaller companies (between 1 and 100 employees) and 102 were employed in large companies. 71\% stated their job role as software developers, 22\% as team lead or product owner, 4\% as operations engineers, and the remaining 3\% listed software architect, researcher or chief technology officer (CTO). The average professional development experience per participant was 9 years ($\pm$ 5).

\textbf{Analysis.} 
We analyzed the distributions of Likert responses and present the results along with the findings from the interview study phase. Furthermore, we examined the responses of three free-text questions on overall differences in development in cloud versus non-cloud environments, restrictions in the cloud, and tooling. Based on these results, we were able to enhance our understanding of some of our findings. %In some cases, we gained new insights that led to the formulation of further findings.

\paragraph{Qualitative Deep-Dive Interviews (Interview$_2$)}
\textbf{Protocol.} Through the examination of open-ended survey questions, we identified 4 topics of high interest: \emph{(1) fault localization, (2) monitoring and performance troubleshooting, (3) cost of operation,} and \emph{(4) design for scalability}. In order to get more profound insights into these topics, we defined an overall set of 27 questions for these 4 topics and conducted another round of semi-structured interviews. We followed the same protocol as in the first round of qualitative interviews. Interviews lasted between 30 and 45 minutes and were audio-recorded.

\textbf{Participants.}
Interviewees were recruited through our personal network. Overall, we recruited 9 participants (D1 to D9), 8 male and 1 female software developer with an average professional development experience of 8 years ($\pm$ 6) from 6 different countries and two continents. All participants were from different companies. 3 participants deployed on PaaS, 5 on IaaS and one on IaaS but also on PaaS.

\textbf{Analysis.} After the interviews and based on the topics and categories identified in the previous two steps, we used open coding to categorize interview statements and gather more profound insights into the difference between cloud and non-cloud development.
%All observations are supported by multiple statements of the interview participants.

%\tf{The following is a summary of the results, not the way you analyzed the data! Please change, i.e. state what you did to analyse the results} Statistical analysis of the responses of our survey led to the validation of 8 of our interview observations. 2 observations had to be discarded based on the survey data. Further, we were able to enhance our understanding of some of the observations, based on free-text feedback given in the survey by a subset of the participants. %\textbf{TODO:} this should be formulated much more exact.
% !TEX root = ../cloud-dev.tex
\section{Findings}
\label{sec:findings}
In the following, we present the findings of our study. We first give a high-level overview of our findings in Section~\ref{sec:overview}, and then provide more detailed results relating to \textbf{RQ1} (Section~\ref{sec:rq1}) and \textbf{RQ2} (Section~\ref{sec:rq2}). 

\subsection{Overview}
\label{sec:overview}

\begin{figure}[h!] \centering
	\includegraphics[width=\columnwidth]{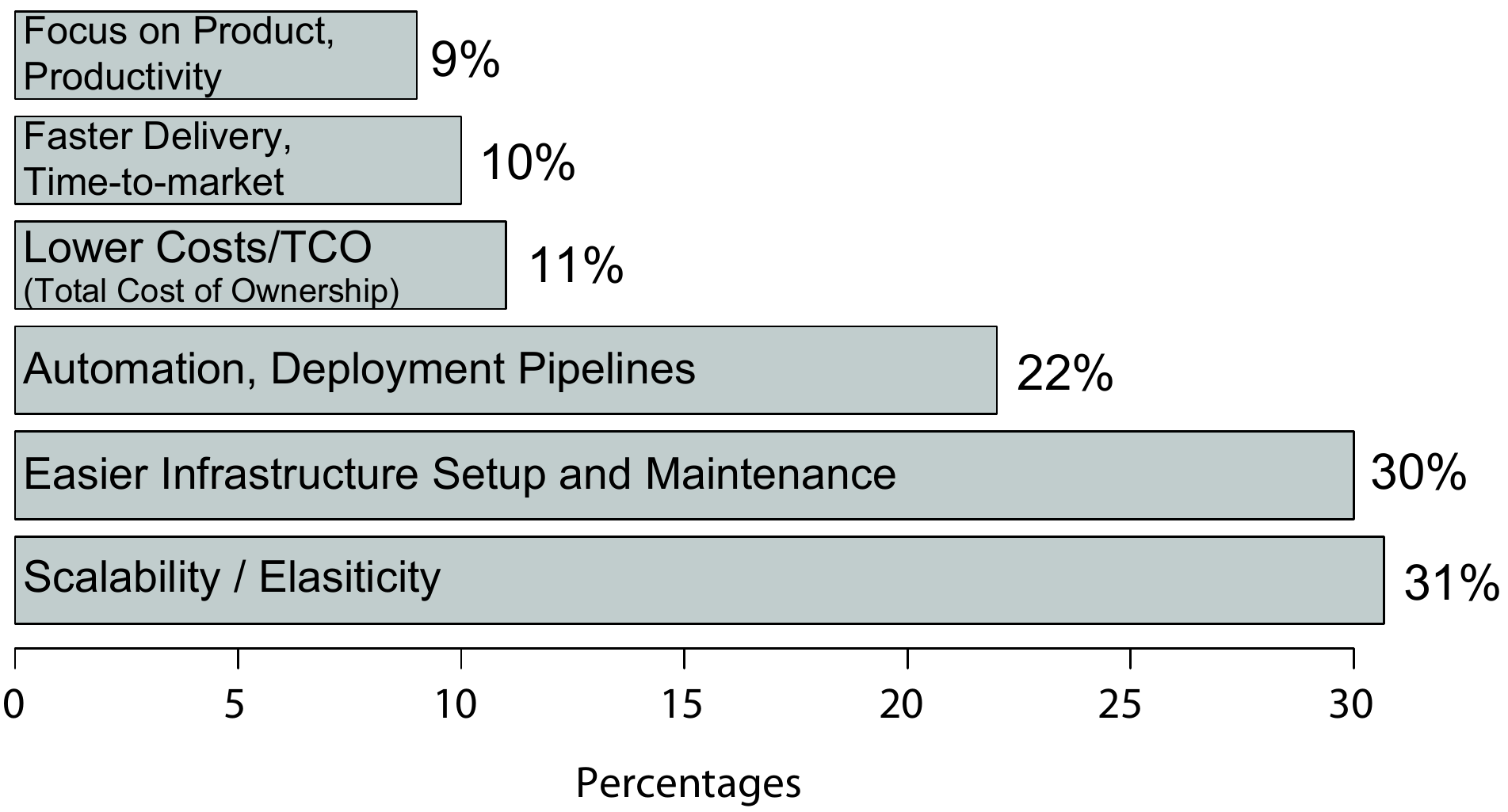}
	\caption{Main Differences in Cloud Development} 
	\label{fig:differences}
\end{figure}

For our study, we were primarily interested in examining what makes software development for the cloud ``unique'', i.e., what differs in terms of processes, tools, and implementation choices, to other development projects. However, early on in our interviews, it became clear that ``the'' cloud does not exist for practitioners. Hence, we asked our survey respondents to list the main characteristics that define cloud computing for them and gathered answers from 160 developers (multiple answers were allowed). We categorized and quantified the most common themes, and present them in Figure \ref{fig:differences}. 
Note that the last three entries in the chart---automation, ease of infrastructure maintenance, and elasticity---are all strongly related, as are the first entries in the chart---focus on product, faster time-to-market.

\begin{figure*}[t]
	\captionsetup{justification=centering} 
	\centering
	\includegraphics[width=\textwidth]{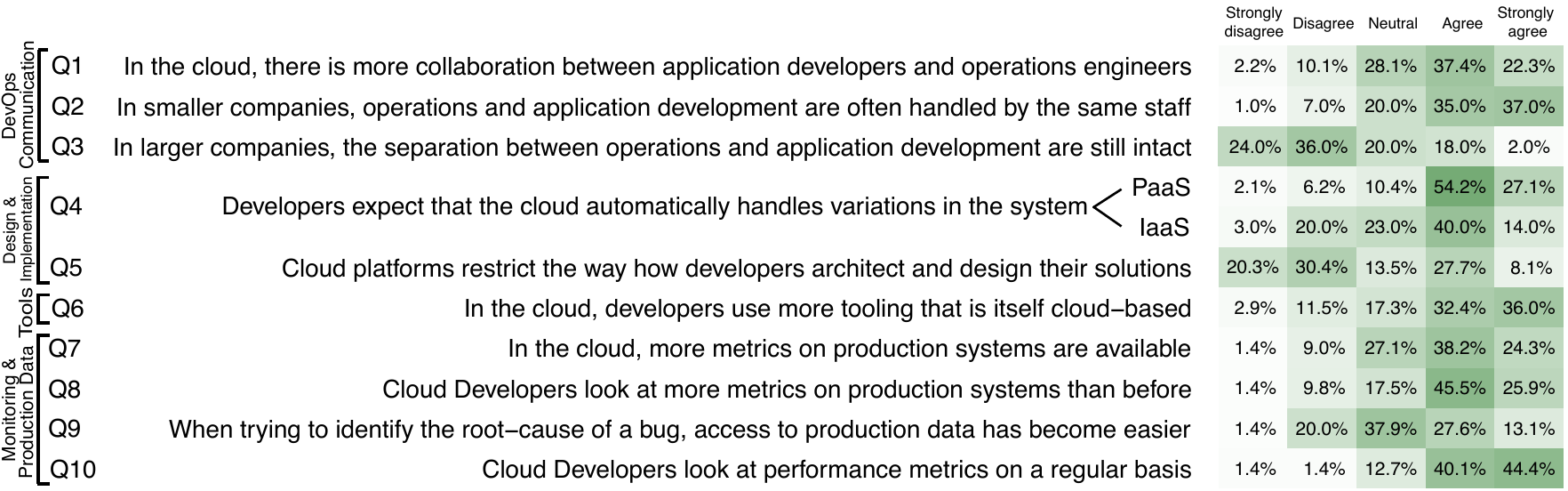}
	\caption{Results from our quantitative survey} 
	\label{fig:findings}
\end{figure*}

The majority of developers thinks about the cloud mostly as a deployment and hosting technology, following either the IaaS or PaaS model. For
these developers, the ability to easily scale applications and the ease of infrastructure maintenance is what makes the cloud unique.  While productivity and faster time-to-market has been named as a distinguishing feature in our interviews, these were only relevant to about every tenth survey respondent. Finally, it is interesting to note that reduced Total Costs of Ownership (TCO) are also only a distinguishing factor for a minority of cloud developers.

Our interviews further revealed that  there is a large mindset gap between developers who think of the cloud as either IaaS or PaaS, and those who think of it mostly in terms of SaaS. For developers where cloud is seen more as a delivery model (i.e., SaaS), how the application is actually hosted tends to be opaque in cloud and non-cloud environments alike and not much changed when adopting cloud computing.
%\interviewquote{I am an application developer, I always have some layer of abstraction below me (\ldots) I was always abstracted from the hardware.}{P8$_{PaaS}$}
%\tf{Since you stated that P1 to P16 were PaaS or IaaS devs in the method section, I'm not sure how P8 is on the SaaS level, maybe you might just have to adapt the table and the participants section.}
%\tf{On a different note, I think it would be good to provide a subscript/superscript showing what kind of developer it is, e.g. P8$_{SaaS}$}

This was, however, different for interview participants working on IaaS and PaaS services. These participants consistently commented on a range of differences. The analysis of our interview transcripts and the survey data shows that the changes when adopting cloud computing for IaaS and PaaS developers fall into the following broad categories: changes to how applications are provisioned and deployed (\textbf{Deployment \& Automation}), changes in how applications are actually built (\textbf{Design \& Implementation}), changes in how problems can be debugged (\textbf{Troubleshooting \& Maintenance}), and cultural changes (\textbf{DevOps Communication}). 

We were also interested in investigating the role of data and operational metrics, as it plays an important role in software development\cite{infneeds}. Indeed, the analysis of our study showed that it plays a major role in cloud development.
%Further, we have seen in the analysis of our study that data and operational metrics play a central role in cloud software development.

 Concretely, the usage of cloud-based tooling has increased (\textbf{Cloud Tooling}), more and more types of data are available for teams (\textbf{Monitoring \& Production Data}), but that developers currently struggle to fully utilize this additional data (\textbf{Usage of Runtime Data}). In the following, we will discuss our detailed findings based on these broad groups of changes.

%\tf{how did you see this "further we have seen that data and ..."? are these some of the most commonly mentioned changes, ...?} 
%\tf{Why does this list not overlap with Figure 3, it seems that some of the top four categories are not int there (e.g. design and implementation or DevOps communication)? Please state this more explicitly.}

In Figure \ref{fig:findings}, we summarize the quantitative results for the Likert-type scale questions from our survey that relate to our core research questions. All detailed results of the survey can be found on our website\footnote{http://www.ifi.uzh.ch/seal/people/cito/cloud-developer-study.html}. We grouped the results in Figure \ref{fig:findings} according to our subtopics. The insights from \emph{Deployment \& Automation} and \emph{Troubleshooting \& Maintenance} resulted primarily from our interviews, rather than the survey.

%For one of the results in Figure \ref{fig:findings}, we were interested specifically in results from respondents working on IaaS or PaaS platforms. We therefore divided the analysis and indicate it accordingly. %Due to survey length concerns, we selected these particular 10 mainfindings from the interviews for the survey. 
In general, we saw a high level of agreement with our interview findings. However, a few individual questions showed some disagreement as well, requiring more detailed study. These aspects are discussed in more detail in the remainder of the paper.

In the following we present the main themes of our study. Where applicable, we provide quantitative results from our survey in the presentation of our findings.

%\tf{are there relevant and not relevant questions? why would you ask non relevant ones?}

%\tf{Not sure what you mean by "individual" questions, please explain. Do you mean a few questions...?} \tf{Are you not discussing Figure 2 a bit? It seems a bit rough to not point out any of these, or will they be picked up in 5.2 or 5.3?}

%\tf{I get the beginning of this overview section, but towards the end I am a bit dazzled what you really want to get across and how you are coming up with some of these points. I was somewhat expecting that figure 3 would show all important aspects, but now it seems like there are some from there, but also a lot others that are not really motivated for why they are there.}
%Where applicable, we provide quantitative results from our survey, where we will reference this Figure.
% Due to concerns regarding the survey length, we selected these particular 13 main observations to be presented to survey respondents.
%   We complement the findings with a selection of quotes from our interviews to deepen understanding. As a part of our interviews had been conducted in German, quotes have been translated if necessary. 

%We conclude each topic with a brief discussion of our results.

% \input{src/definition_cloud}
% 
% \input{src/difference_cloud}

%\input{src/underlying_cause}

% !TEX root = ../cloud-dev.tex
% TODO: add how releases have evolved, with division in company size

% \subsection{Process, Communication, and Team Dynamics}

\subsection{Application Development and Operation}
\label{sec:rq1}

In this section, we report on how cloud computing has affected application development
and operation and on the main drivers---API-driven infrastructure-at-scale and cloud instance volatility---for these changes.
%We relate our findings to a standard development process (based on \cite{[sommerville:software_engineering]}) in the following. 
%(1) Project Management Methodology and Release Management, (2) Design \& Architecture, (3) Implemention, Integration \& Testing, (4) Deployment \& Automation, and (5) Troubleshooting \& Maintenance.

\subsubsection{Deployment \& Automation}
\label{sec:dep}
 
Our interviews have shown that elasticity, ease of infrastructure maintenance, and automation can be broken
down into two fundamental aspects that drive most changes of software development in the cloud:
(1) \emph{API-driven infrastructure-at-scale} and (2) \emph{volatility of cloud instances}. Both these aspects have ripple effects on almost every aspect of cloud development.

%\tf{where have they shown this. If that's a finding, than state it as such.}
%Hence, we will often refer back to these major changes when discussing
%other aspects of cloud development.

%We will now briefly explain these two causes and, where applicable, refer to them when presenting our findings in the following sections.

\subheading{API-driven Infrastructure-at-Scale} Infrastructure-at-scale refers to the
ability to quickly spawn up (and discard) many compute instances using an API. This ability requires
more automation on many levels, including infrastructure, environment and test.
In \emph{IaaS clouds}, automation happens by defining your infrastructure as a
set of software artifacts (see IaC in Section \ref{sec:bg}). This allows for automatic provisioning of
newly created instances for different purposes (e.g., scaling up, setting up a
test environment). In {PaaS clouds}, applications are required to be packaged in
a way to be easily reproducible (e.g., containers, buildpacks) to manage this
automation behind the scenes. All interviewees deploying on IaaS agreed that the
use of IaC tools (e.g., Chef, Puppet) or other means of automation (e..g, shell scripts) for all automated provisioning of their
infrastructure has become essential in the cloud.

\subheading{Volatility of Cloud Instances} Cloud instances can be started up and
shut down for various reasons. This volatility happens either through (1) the
cloud provider shutting down instances, (2)  the load balancer spawning or
shutting down instances, and (3) the application itself shutting down a
dynamically allocated instance that finished its work. %\tf{is spawning commonly used for the cloud context?}

One implication of instance volatility is that all infrastructure definition and server configuration has to be implemented in code. If provisioning and configuration is not automated, it is bound to be lost when instances are discarded. This means that every change in infrastructure results in a new deployment of the system.
 Four of the interviewees that deploy in IaaS referred to this practice as \emph{immutable infrastructure}:
\interviewquote{We have now moved to strict Immutable Infrastructure in our deployment. We don't even put SSH keys into instances anymore, making changes to existing infrastructure impossible}{P12$_{IaaS}$}
In PaaS, the infrastructure is managed by the cloud provider. Hence, the infrastructure is immutable for developers by default.

\subheading{Infrastructure Transparency} All interviewees deploying on IaaS mentioned the use of either IaC tools
or shell scripts for all automated provisioning of their infrastructure. 
They argue that this brings them transparency regarding their infrastructure:
\interviewquote{What happens in our infrastructure is a lot more obvious. Everything we do on that level [infrastructure] is over code  (\ldots) So, I don't need ask my colleague what he did to get that process running - I just look at the code and maybe the commit history}{P9$_{IaaS}$}

\subheading{Virtual Containers for Automation} Furthermore, three interviewees describe a push in virtualization from virtual machines towards virtual containers (e.g., LXC\footnote{\url{https://linuxcontainers.org/}}, Docker\footnote{\url{https://www.docker.com/}}) for their automation:
\interviewquote{Virtual machines are too slow in a large scale (\ldots) Speed matters, also in integration testing. When I can make a build take 3 minutes instead of 20, that's a huge win}{P13$_{IaaS}$} 
%\tf{3 interviewees are very few, should this definitely be in here, i.e. this whole point?}

Traditional virtual machines impose a large performance overhead due to the additional virtualization layer. Containers allow for much faster start-up times and are, therefore, also increasingly used as a base for PaaS~\cite{tang:14}.

%Three of the interviewees used the term \emph{Immutable Infrastructure} that refers to .  
%In turn the need for automating infrastructure in this way has implications on the deployment process. 

\summary{Infrastructure provisioning and application deployment in the cloud are largely automated. Servers
are not seen as durable entities. Hence, any changes to infrastructure need to be defined in code.
This also leads to more transparency concerning infrastructure changes.}

\subsubsection{Design \& Implementation}
\label{sec:design}

%buy into the cloud ecosystem and face restrictions. we asked what kind of restrictions they face
% 
% We found that developers face different kind of challenges, depending on the cloud model they deploy to. Thus, we further divide this section to address challenges developers have in either \emph{IaaS} or \emph{PaaS}.

%\tf{you have a motivation / introduction for the section before (5.2.1) but not in here, why? maybe either take out the other one or add one here; probably adding one would be good and providing a motivation for the points you will be making in the following (e.g. since they were the most frequent,...), not just saying you will discuss them. }
In this section, we report on how restrictions in the cloud influence how developers design and implement their applications.

\begin{figure}[h!] \centering
	\includegraphics[width=\columnwidth]{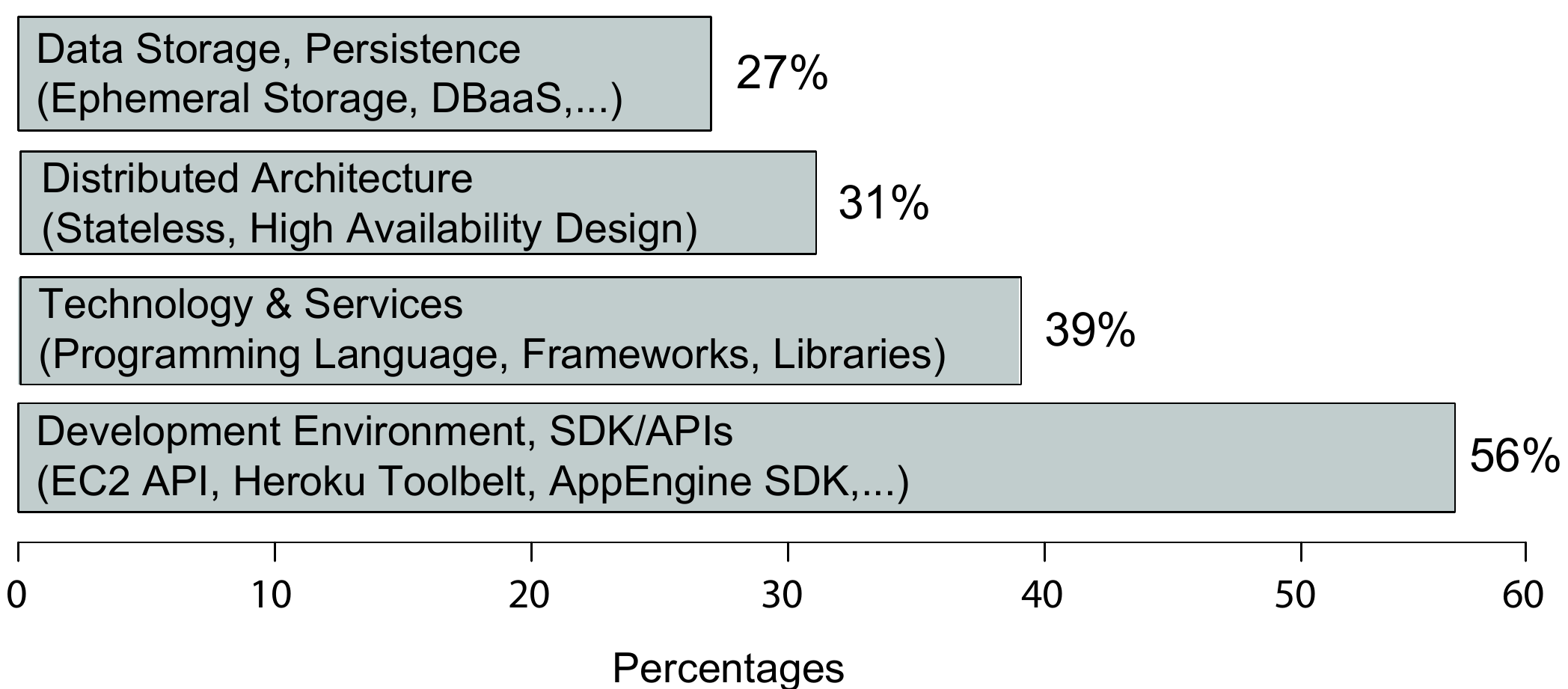}
	\caption{Most significant restrictions developers have encountered on cloud platforms} 
	\label{fig:restrictions}
\end{figure}

\subheading{Design Restrictions} As part of our survey, we asked whether the survey respondents face limitations in application architecture and design specific to the cloud. Results to this question were  non-conclusive, with 51\% of respondents disagreeing and 36\% agreeing (see Q5 in Figure~\ref{fig:findings}). However, 119 respondents have still (in a free-form field in the survey) stated multiple restrictions, which we categorized and quantified in Figure \ref{fig:restrictions}.

It is not surprising that there are technical restrictions regarding the supported SDKs, libraries, and frameworks.
However, to our interview participants, these restrictions were not all negative.
Some developers feel that technological restrictions allow them to focus more on delivering business value, instead of tinkering with low-level technology choices:

\interviewquote{But I don't wanna make those decisions [on technology] anyway. (...) The cool thing - from a product design point of view - is, that you know what works on AppEngine and what doesn't}{P11$_{PaaS}$}

More interesting is that close to a third of all respondents also consider the cloud to restrict
them in the way they actually architect and design their applications. 
These restrictions in design and architecture are primarily caused by API-driven infrastructure-at-scale and volatility of cloud instances,
as discussed in Section~\ref{sec:dep}.

\subheading{Design for Failure} Volatility of cloud instances naturally forces
developers to build highly fault-tolerant applications, as IaaS providers reserve
the right to shut down any resources at any time, on short or without any prior
notice:

\interviewquote{One interesting thing that is very cloud specific and influenced
our architecture is, that the cloud provider tells you, we can kill your machine
any time we want.}{P9$_{IaaS}$} 

Well-known cloud users have already adopted this mindset. Netflix,
for instance, has stated that they use an application called Chaos Monkey\footnote{\url{http://techblog.netflix.com/2012/07/chaos-monkey-released-into-wild.html}} to randomly terminate cloud instances in production,
to force application design that can tolerate such failures when they happen unintentionally.

%by the provider.

% \paragraph{Higher Expectations} In PaaS, developers expect the platform to handle certain operational aspects in combination with the software stack for them. Operational aspects include, for instance, variations in system load and fail-over/recovery of services and the underlying infrastructure.  What aspects the software stack handles differs among platform providers, but, generally speaking, includes a variety of middleware services that can be accessed through the platform's API.  A recurring theme among interviewees developing on PaaS was that they do not care about system load, as they expect the platform to take care of scalability issues and provide elasticity out of the box. 
% 
% In our survey we asked whether the participants expect the cloud to handle load variations for them. 81.3\% of PaaS and 54\% of IaaS developers agreed or strongly agreed with that (see Figure \ref{fig:findings}). Surprisingly, even on IaaS, 54\% agree that the cloud automatically handles variations in system load for them, although this is not an inherent feature of IaaS solutions. Developers agreeing with this statement might refer to the ease of acquiring server instances that enables the handling of increasing system load through horizontal scaling. Furthermore, IaaS vendors sometimes offer dedicated load balancing instances (e.g., Amazon's Elastic Load Balancer) that facilitate these capabilities.

\subheading{Design for Scalability} Scalability is the most named difference in cloud development for survey participants (see Figure \ref{fig:differences}). In our deep-dive interviews we asked participants to explain how scalability considerations influenced them during design and implementation.

All interviewees stated that they always have scalability in mind, even when designing very simple cloud applications:
\interviewquote{Even if the customer needs only 1 server, we have an ELB (Elastic Load Balancer) anyway, because we expect everything to grow}{D3$_{IaaS}$}

%The ease of acquiring compute instances has made the life of operators easier. Software developers, on the other hand, now need to adhere to different design (i.e., statelessness, using backing services, etc.).
The Twelve-Factor app\footnote{\url{http://12factor.net/}} design has become the de-facto standard when it comes to best practices when building cloud applications. A few of the interview and survey participants referred particularly to this manifesto, while others often referenced similar practices for implementing scalability under a different name.

An alternative approach to implement fault-tolerant and scalable cloud applications that was mentioned are microservices~\cite{newmann:15}: 

\interviewquote{We have divided our application into many services. For one specific service we kill off and start new instances all the time, also to have proper redundancy.}{P9$_{IaaS}$}

However, in our interviews, only five participants mentioned either currently using or having plans to move to a more service-oriented system design in the near future.

%Has an impact on capacity planning (there is other scientific work on that - cite?)
%Infrastructure at Scale: Getting new hardware in the cloud is easy, ELBs come configured and ready
%SW Dev has changed, in order to scale, we need to adhere to a different design (stateless, using backing services, etc.) —> more details in next section
%"12factor app” is de facto standard when implementing - all interview partners mentioned either 12factor or mentioned techniques that follow the same principles
%Important one: Stateless instances, i.e. can’t store data on the instance itself, needs backing service that is often also a cloud service (e.g., Database as a Service, File storage as a Service)

%workflow changed (looking at metrics, will be covered in the next section where we cover tools and data)

%local environment - cloud environment mismatch

\summary{The cloud imposes some restrictions on how applications can be built, in technological and software
architectural terms. Specifically, applications need to be designed for
scalability and fault tolerance. These restrictions are also seen as positive as they enforce
the best practices and foster business value.}

\subsubsection{Troubleshooting \& Maintenance}
\label{sec:troubleshooting}

Activities that happen after the application has been deployed (i.e., troubleshooting and maintenance) have probably seen the biggest change in cloud development. Servers are not seen as durable entities anymore. Therefore, infrastructure maintenance has become an activity that now has to deal with adapting infrastructure code files rather than tweaking on live server instances.

\subheading{Fault Localization} Fault localization, the activity of discovering the exact locations of program faults, cannot be done through ad-hoc inspection on, for instance, log files, memory dumps, or system metrics on live production servers anymore. 

This means that every information that is of interest in the maintenance phase of the development life cycle must be specified before deployment and collected at a central repository, otherwise information runs the risk of being lost due to cloud instance volatility. These techniques are not necessarily bound to the cloud.
%, they can and should be used in standard distributed architectures as well.
However, in the cloud these best practices are seen as mandatory:

\interviewquote{In the cloud you are forced to use best practices, you are forced to use automation. You are forced on not relying to having root access to jump onto a machine and search logs by hand. You are forced to use better practices like aggregation.}{D6$_{PaaS}$}

% other quote: Difference is that in all the system you can get away with doing all the things by hand, in the cloud it’s simply not possible to do that

%This development might suggest a tighter collaboration between developers building functional features and operations engineers taking care of production aspects. We investigated a bit further on how the relationship between these two entities has evolved.

\subheading{Reproducing Issues} In terms of reproducibility of  issues in a local environment for fault localization,
our interviewees were somewhat divided. 
On the one hand, this task has become more difficult, as cloud applications are inherently distributed and reproducing a distributed environment locally is generally a difficult task.
On the other hand, deployment automation makes it easier and faster to spin up a staging or testing environment in the cloud to reproduce issues. However, our interview partners also mentioned the additional cost involved of spinning up environments as a reason for first attempting to reproduce issues locally.
Some interviewees also stated that proper end-to-end monitoring\cite{cito:14} and request tracing needs to be in place to be able to reproduce issues correctly: 
\interviewquote{If you missed some logs and the instance is already gone, have fun reproducing your environment}{D8$_{PaaS}$}
Especially with public cloud providers, hardware characteristics also need to be tracked, as you never know what specific hardware configuration will be served.

%- missing that data has to be replicated too, i.e. knowing what the database state was e.g.

%Volatility also goes down on instance level: Can I reproduce the issue? Needs tracking of exactly what hardware you received from (public) cloud provider
%Statement: "if you missed some logs and the instance is gone, have fun reproducing it" —> requires to plan metrics and what to log way beforehand
%Troubleshooting also needs request tracing to identify the root-cause of issues in a distributed system

\summary{Troubleshooting has changed, as problematic cloud instances are often not accessible or already discarded, rendering hot fixes or searching for logs in production impossible. Instead, relevant logs must be defined beforehand and collected in a central repository. These best practices are not exclusive to the cloud, but instance volatility makes their usage in cloud computing mandatory.}

\subsubsection{DevOps Communication}

As already discussed in Section~\ref{sec:bg}, it is often argued that cloud computing goes hand in hand with a DevOps style of communication, which leads to higher collaboration between software developers and operations engineers. In our interviews, this notion was not undisputed. While 12 of the interviewees have agreed with the DevOps vision, the remaining participants (mostly from enterprise companies) have argued that there are still dedicated development and operations teams that more or less work as silos. Even companies that self-identified as following a DevOps approach still seem to generally have a separation between engineers that solely implement functional features, and engineers that mostly develop infrastructure code:
\interviewquote{We have our server/DevOps guy. (\ldots) He handles the whole monitoring and tools thing}{P9$_{IaaS}$}

%While five interviewees (P9-P13, D2-D6, D8, D9) have agreed with the DevOps vision, the remaining participants (including most of our study participants from larger companies) have argued that there are still dedicated development and operations teams, and that there is still considerable friction between those teams. 

\begin{figure}[h!] \centering
	\includegraphics[width=\columnwidth]{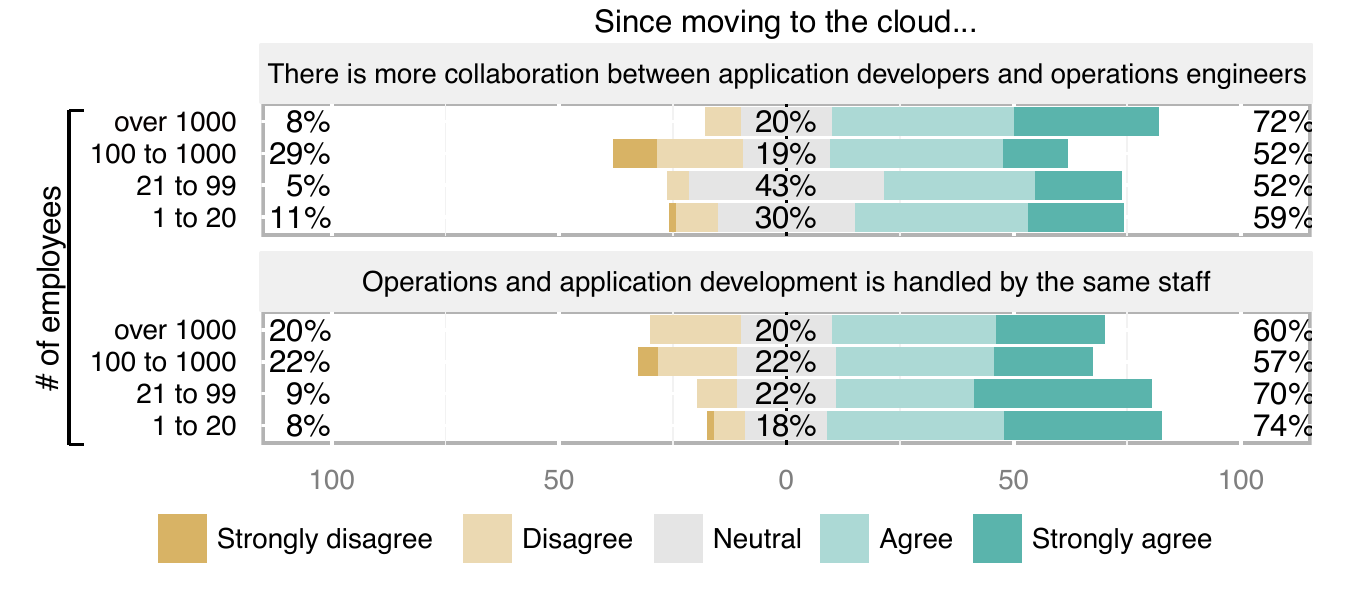}
	\caption{Results regarding Team Communication grouped by company size} 
	\label{fig:survey_communication}
	\vspace{-1em}
\end{figure}

In our survey, there was a general agreement with these observations (see Figure~\ref{fig:survey_communication}). The survey responses show that, especially in large companies, communication and interaction has increased between application developers and operations engineers (72\%). 
Contrary to our interview study, the difference between the responses of developers working in smaller or larger companies in terms of collaboration between development and operations is not large. For smaller companies, 70-74\% tend to agree that their operations and development are now handled by the same staff, versus 57-60\% in larger companies. This data suggests that, even in larger companies, the gap between development and operations activities converges. 

\summary{Close to 60\% of developers across companies of all sizes build applications following the DevOps notion of converging development and operations teams. However, even in a DevOps team, there are still dedicated engineers that are responsible for maintaining the infrastructure code.}

%Based on our interviews and survey data, it seems that the notion of DevOps is indeed becoming a reality. Between half and 3 out of 4 developers state that cloud computing has brought them closer together with their operations team, and especially smaller companies are actively adopting the DevOps terminology and tooling (including IaC). However, quite contrary to the purist DevOps idea, our interviews have also shown that even small companies that self-identify as having a DevOps approach, still often think in silos of software developers and operations engineers. The main difference is that now these operators are referred to as ``DevOps engineers''. These engineers are rarely responsible for building functional features. Instead, their job is often to monitor performance (as discussed in Section~\ref{sec:data}) and maintain the IaC code for provisioning

%Who defines metrics and logs?
%Logs as communication tool with operations

%\input{src/communication}

%\input{src/process_communication}

%\input{src/impact_development}
% !TEX root = ../cloud-dev.tex
% \subsection{Tools and Data}

\subsection{Changes in Tools and Metrics Usage}
\label{sec:rq2}

In our research, we were particularly interested in how the usage
of tools and production data has changed in cloud development projects. We report on (1) how tooling has evolved in the cloud, (2) what kind of metrics are available now, and (3) how these metrics are utilized.

% !TEX root = ../cloud-dev.tex
\subsubsection{Cloud Tooling}
\label{sec:tooling}

%In this section, we report on the tooling that is utilized for developing applications for the cloud.

\begin{figure}[h!] \centering
	\includegraphics[width=\columnwidth]{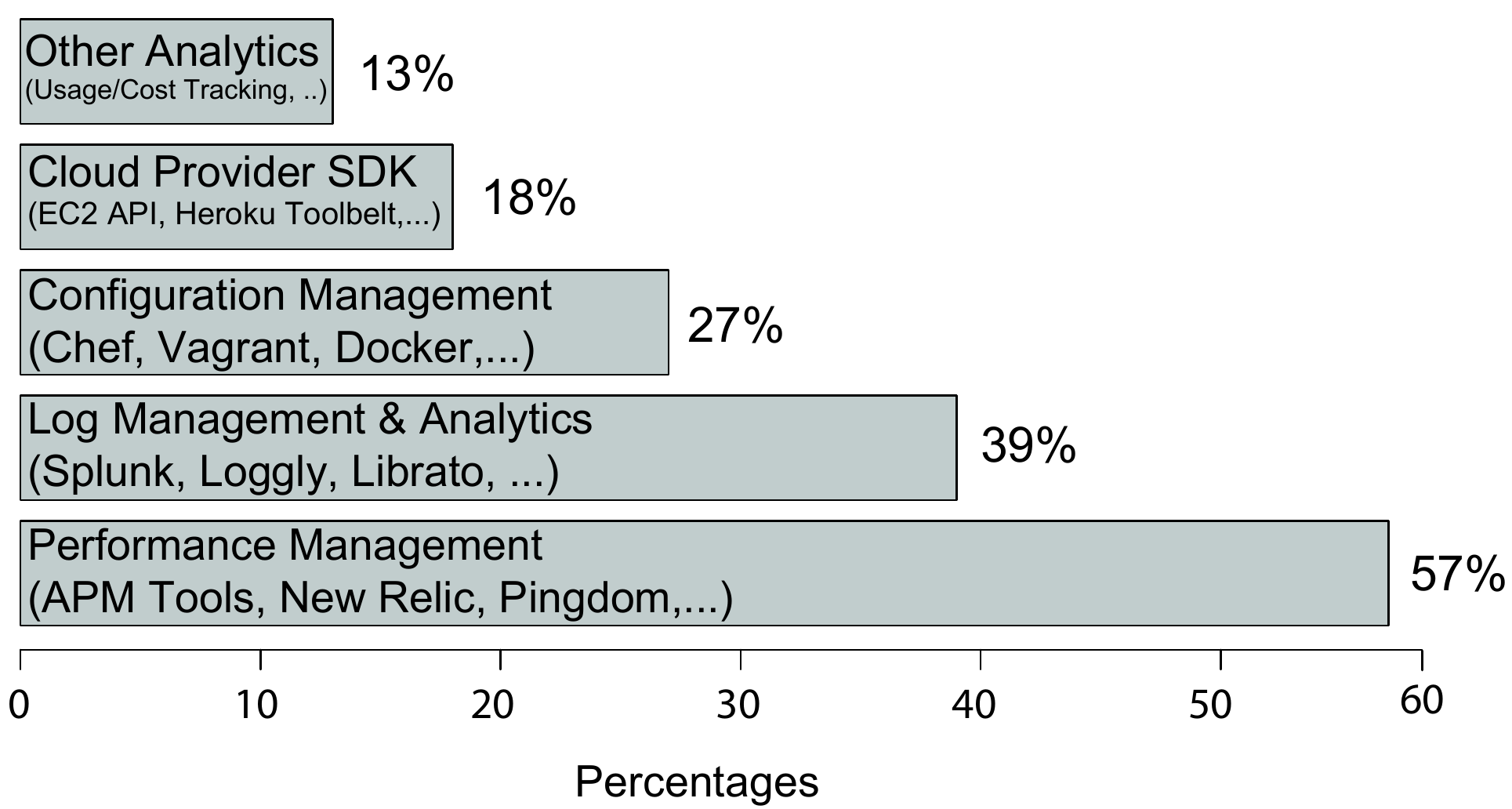}
	\caption{Tools used specifically for cloud development} 
	\label{fig:tools}
	\vspace{-1em}
\end{figure}

\subheading{Tooling for the Cloud} In the survey, we asked what tools developers specifically use in development for the cloud, which they have not used before. 124 people responded to the question with multiple tools, which we categorized and quantified in Figure \ref{fig:tools}. We also asked whether cloud-based tool usage has generally increased. 67\% of respondents agreed with this statement, while 15\% disagreed (see Q6 in Figure \ref{fig:findings}).

\emph{Performance and Log management} top the list of tools survey respondents use specifically for cloud development.
This can again be attributed to the notions of API-driven infrastructure-at-scale and cloud instance volatility, which we have already discussed extensively in Section \ref{sec:rq1}. 

\subheading{Tooling in the Cloud}  We also observe an increase in the usage of tooling that is itself cloud-based. However, we cannot differentiate whether this rise has to do with development on cloud platforms, or with the fact that more and more tooling is moving to a SaaS model in general. Interviewees agreed that, when building cloud applications, many of the tools previously installed in the local infrastructure or on the developer machine are now also in the cloud. This includes tools for monitoring, analytics, and configuration management. Four of the interviewees even used an entirely Web-based IDE for most of their development tasks. All others moved at least part of their supporting tooling into the cloud. The interviewees developing entirely in the cloud (i.e., through a cloud-based IDE) expressed the overall relief of not having to maintain or configure a local development environment:% (general setup, plugins and updates, backup):

\interviewquote{I don't need to take care of backups, or updates of the IDE. I can use the same setup everywhere. (...) Also there's no hassle about that missing plugin and that wrong configuration.}{P1$_{PaaS}$}

However, cloud-based IDEs were not mentioned by survey respondents at all as part of their tooling. Hence, we conclude that cloud-based IDEs have, unlike indicated by our interview study, not yet found mainstream adoption.

%\subheading{Deployment Automation Tooling} Another theme that emerged were automation tools in the deployment process. All interviewees deploying on IaaS mentioned the use of IaC tools (see Section \ref{sec:bg}) for all automated provisioning of their infrastructure. 
%The strength of this kind of automation comes through combination with virtualization.  % bessere ueberleitung...

%\interviewquote{Every change on our infrastructure is written as code in either Ansible or Puppet}{P15}
% We interpret this as an indication that these IDEs have not seen a lot of adoption so far.

\summary{Cloud-based tool usage has generally increased. Performance, log management and analytics tools are  most used specifically for cloud development. Many of these tools are themselves cloud-based. However, despite some advantages, cloud-based IDEs have not yet found mainstream usage.}

%Thus, logs are scattered across instances, which increases the difficulty for locating certain logging statements in the debugging process. This also means that identifying root-causes of performance issues, and finding their origin has become increasingly harder, often requiring end-to-end performance monitoring\cite{cito:14}. 
%However, these issues are not new to enterprise companies, which already had to deal with the management of large server clusters in the past. Cloud computing has made these issues more visible to a critical mass of developers. %alternative: to the mainstream of developers

%\resetObservations
%\begin{observations}
%\simpleobservation{In the cloud, developers use more tooling that is itself cloud-based}
%\simpleobservation{E}{Developers use more tools for cloud development that are themselves cloud-based}
%\simpleobservation{E}{Performance-, Log-, and Configuration-Management tools are increasingly and specifically used in cloud development} 
%\end{observations}

% !TEX root = ../cloud-dev.tex

\subsubsection{Monitoring \& Production Data}

\subheading{Metric Awareness}
All interview participants have mentioned using solutions for log aggregation and central operational metric collection for cloud development. The survey supports these claims with performance metric and log management tools being deemed the most important. What we have also observed is that the enforcement of these practices has led to more awareness for metrics:
\interviewquote{Software Developers have not - until recently - seen metrics as very important. They saw that stuff as an operations thing}{D5$_{PaaS}$}

In our survey, 72\% of cloud developers agree that they look at more metrics on production systems than before (see Q8 in Figure \ref{fig:findings}).

%This development might suggest a tighter collaboration between developers building functional features and operations engineers taking care of production aspects. We investigated a bit further on how the relationship between these two entities has evolved.

%\subheading{Data Access}

%\subheading{Organizational Limitations} Five interviewees reported that, when trying to find the root cause of a bug, obtaining production logs has become harder than in non-cloud environments. However, the difficulty in these cases did not arise from technical limitations, but rather from limitations of organizational nature:
%\interviewquote{You need an approval from your manager, then from the customer, who reviews the whole process (...) It's all very conservative}{P6}
%In these particular instances, customers expressed privacy concerns they had with cloud platforms leading to higher rigor in organizational processes to obtain any kind of data from production systems. 

%Access to production data has become easier according to survey respondents. For developers on PaaS systems, access to production systems is probably seen as easier because platform providers aim on offering a holistic solution to cloud developers. In IaaS, the majority of respondents were neutral on the topic. This is because, in this cloud model, access to the systems is unchanged compared to non-cloud environments.
%Our original observation was derived from interviewees within a certain company and probably originated more from specific company policies, rather than being inherent to cloud development.

\subheading{Metric Availability}
In general, interviewees expressed that they now they have a much richer set of monitoring metrics on production systems available:
\interviewquote{Especially now in the cloud (with Heroku or Amazon) they provide so much data. Putting this data in a graphing tool and looking at it once in a while has become increasingly easier.}{P12$_{IaaS}$}
From the survey, we see that 62\% of respondents agree that they actually have more operational metrics available than before (see Q7 in Figure \ref{fig:findings}). When asked whether access to production data has become easier, survey participants were divided: 21\% disagreed, 38\% were neutral and 41\% agreed with the statement (see Q9 in Figure \ref{fig:findings}).

We investigated this claim further and found that metric utilization has increased both in quantity and dimension. In addition to technical system-level metrics (e.g., CPU utilization, cache hits), more teams now look also at business metrics (e.g., customer retention, number of logins) to guide their decisions.
However, system performance metrics are deemed to be the most interesting by our survey participants. 84\% state that they look at performance metrics on a regular basis (see Q10 in Figure~\ref{fig:findings}).

%We were interested in seeing \emph{how} metrics are utilized in the development process and investigated this topic further in the following section.

%\subheading{Business Metrics}

%- we have always looked at system level metrics (cpu, cache hits, etc.)
%- now people look at more metrics and metrics that they look at have expanded (more metric types) -> business metrics
%- more data

\summary{Cloud developers look at more production metrics. In addition to system-level metrics, business metrics are becoming more relevant. However, application performance is most often still measured via system-level metrics.}

\subsubsection{Usage of Runtime Data}

An increase of tooling to acquire data, more metric awareness and availability made us interested in how developers use metrics in their work. Through analysis of our initial interviews and the open questions in our survey, we identified that \emph{performance} and \emph{cost} were the metrics of high interest to cloud developers. In the following we describe how our interview participants utilize this data in their regular development work.

\subheading{Performance Troubleshooting}
%In our deep-dive interviews, we asked the study participants to walk us through a particular performance problem they faced and explain how they utilized data from production systems to solve the problem.
In our deep-dive interviews, we investigated how our study participants approach a particular performance problem by utilizing data from production systems to solve the problem.

%We were able to identify a common thread in the way our interview participants approach these issues.
%The first common theme is that metrics are usually not defined by the developer, but by someone who is concerned with running the application in production (i.e., DevOps engineers\footnote{Depending on who we were talking to, this role was also called operations engineer or system administrator}). Metrics were then all provided in dashboards accessible to everyone in the team. DevOps engineers were the ones that actively observe the data on a regular basis. Developers followed a more reactive approach, i.e., they acted on alerts or reports on the issue tracker.
A common theme is that developers follow a more reactive approach to metrics, i.e., they act on alerts or reports on the issue tracker provided by someone who is concerned with running the application in production (i.e., DevOps engineers, operations engineers, or system administrators).
%\footnote{Depending on who we were talking to, this role was also called operations engineer or system administrator}.).
Metrics are typically provided in dashboards, accessible to everyone in the team, but mostly used by operations. However, when actively debugging a known performance issue, all interview participants would first go \emph{"by intuition"} and reproduce the issue in their own development environment, before inspecting how metrics have evolved in production in the provided dashboards. Only if the local inspection does not yield any results, they would dig deeper into production data. Interviewees stated that they choose to ignore the data at first, because it is rather cumbersome to navigate performance dashboards, while at the same time navigating through code:
\interviewquote{I try to reproduce and solve the issue locally. Looking for the particular issue in the dashboard and jumping back and forth between the code is rather tedious}{D2$_{IaaS}$}

% original quote: “Ich probier das ganze mal lokal zu reproduzieren und zu lösen. In den Dashboards nach dem Problem herumsuchen und immer wieder in den Code zu springen ist schon mühsam”

%When prompted on why they choose to ignore the data at first, they stated that it was rather cumbersome to navigate their performance dashboards, while at the same time navigating through code:

%intution vs. data driven

\subheading{Costs of Deployment} 
To most of our interviewees, the costs of deployment in the cloud were generally deemed as important. 
%In our survey, 56\% agree that they look at costs on a regular basis (see Q11 in Figure \ref{fig:findings}). 
However, developers seem to only in the abstract be aware that their design and implementation decisions have an influence on deployment costs. When confronted on how the costs of deployment (especially of specific code changes) are used in their daily work, it became clear that costs are not a tangible factor for developers. Some interviewees argued that having this information more wide-spread and accessible would be interesting to them, but does not fall into their responsibilities:
\interviewquote{I have no idea about the costs. I can just read it in the logs sometimes that in production we spawned 20-30 servers. It would be interesting to know, but it's not really important for application development}{P5$_{PaaS}$}

Not developers, but software architects or the CTO are concerned with the overall costs of operation. However, even for these roles, costs were considered in a post-design phase and currently do not influence their design decisions directly.

\summary{Currently, developers struggle to use the abundance of available runtime metrics. Rather, they often solve performance problems "by intuition" in a local environment. More detailed inspection of metrics is only used when this approach does not lead to a solution. Cloud costs are deemed as important, but are not tangible to developers.}

\section{Discussion}
\label{sec:implications}

%Main Insights:
%\begin{itemize}
%	\item Infrastructure-at-scale and cloud volatility are drivers for major changes in cloud development
%	\item Overall, the resulting techniques are best practices that can be also be utilized in non-cloud environments (i.e., log aggregation, request tracing). However, in cloud development you are forced to use these best practices.
%	\item This has changed the need for automation, the way we troubleshoot
%	\item It has also changed the way we architect our applications. Cloud imposes restrictions on design and implementation to support scalability.
%	\item More metrics are available in the cloud, both in quantity and in dimension (i.e., people now use business metrics)
%	\item However, developers often go by intuition when making decisions because many metrics are not actionable to them
%\end{itemize}

We now present, in a condensed form, the implications of our study results
for practitioners, as well as the main challenges that cloud developers
face. These form open problems that academic research and vendors
of cloud-related tooling need to address to improve the experience
of developers.

\subsection{Implications for Practitioners}
\label{sec:impl}

Our study has shown that there are a number of best-practices for building useful applications on top of IaaS and PaaS systems. These practices are not necessarily bound only to cloud development, but the nature of cloud infrastructures make these practices central for successfully deploying in a cloud. In the following we present a set of guidelines for software development in the cloud resulting from the findings of our study. %\tf{this sentence and the one before after the section 6 heading can be shortened}
 
\textbf{Cloud instances are \emph{volatile}. Never assume that any instance will
exist forever.} In both, IaaS and PaaS cloud systems, backend instances come and go.
Logs, configuration changes, or hot fixes stored only on a cloud instance are bound to
be lost. As we have seen in Section \ref{sec:dep}, cloud developers should treat cloud instances as \emph{immutable} black boxes,
which they in general cannot fix, and in some cases cannot even log into. This requires
a change in mindset for engineers used to having full control over their infrastructure. %\tf{I really like these findings and this whole section. However, I think you need to reference/relate back to your previous findings a bit more if possible, either by directly referencing your findings (you could call them finding F1, F2, .. and then just use these letters in here, or shortly state it in here.}

%\tf{These statements are very long and it seems like you have a ton of unrelated findings in the end with the boxes earlier on, the figures as well as the bold statements in the implications. Try and relate them if possible and shorten them here. Maybe a whole table that maps the findings from the studies to the implications would be good, but it depends a bit on the mapping, i.e. if it is easy to make or not.}

\textbf{\emph{Anticipate runtime problems} and define relevant logs and
operational metrics for fault localization before deployment. Use log management
tools to centrally collect this data.} Related to the volatility of cloud
instances, developers need to start thinking about how to localize and debug
runtime faults already during development. This includes defining useful
debugging statements, logs and operational metrics prior to deployment, as well
as setting up and using tools to centrally \emph{collect, persist, and analyse}
these metrics outside of volatile instances. As discussed in Section \ref{sec:troubleshooting}, cloud instances are black boxes,
drilling into unanticipated problems after deployment is often impossible.
Hence, cloud developers should be aggressive in what they log and what
operational metrics they trace. It is easier to filter out data that turns out to be
unnecessary than to debug problems for which the relevant logs and metrics have
not been collected.

\textbf{Scalability and fault tolerance need to be \emph{first-class citizens} in application design.}
While most distributed and Web-based applications have historically striven to be scalable and fault
tolerant, at least to some degree, these concepts have become even more central in cloud projects. API-driven
infrastructure-at-scale means that essentially any application or component can scale
up dynamically (e.g., to react to increased load). Instance volatility means that
runtime faults are bound to happen. As such, scalability and fault tolerance need to be
first-class citizens when designing applications (see Section \ref{sec:design}). \emph{Best-practices for cloud
application design} (e.g., the 12-Factor App) take this into account and should not be
compromised by cloud developers. PaaS systems often enforce such application
design through restrictions (e.g., in terms of statelessness). Developers should not
aim to circumvent, but rather embrace those restrictions.   

\textbf{For IaaS, \emph{automate} server provisioning and configuration, for instance using IaC tools.}
API-driven infrastructure-at-scale requires that the process of instance provisioning and
configuration is fully automated. Besides being able to scale up quickly, this also has
the additional advantage that knowledge about how servers are configured is explicitly
documented in scripts or IaC code, and versioned in the project's version control system.
This allows developers to \emph{revert}, for instance, erroneous changes in the configuration of a cloud instance
just like they would revert a broken application code change. It also makes infrastructure configuration and evolution 
explicit for other developers and DevOps engineers leading to more \emph{transparency}, as discussed in Section \ref{sec:dep}.
While some cloud developers currently use scripting (e.g., \texttt{bash})
for this purpose, the usage of dedicated IaC tools has additional advantages. Most importantly,
IaC allows for reuse of existing open source provisioning and configuration code, and supports
unit testing well. 

\textbf{Embrace the \emph{tools and data} the cloud provides you.} As elaborated in Section \ref{sec:rq2}, we have seen that cloud developers have access to more tools and data. However, we have also seen that many developers still rather go \emph{``by intuition''} when debugging problems rather than analyzing
provided operational metrics. We argue that, besides better tooling (see Section~\ref{ref:challenges}),
a change in mindset is required for cloud developers to fully embrace the additional
options for debugging and maintaining applications that the cloud environment
provides them.     

\subsection{Challenges for Cloud Development}
\label{ref:challenges}

In addition to the best-practices and guidelines outlined in Section~\ref{sec:impl}, which
cloud developers can already implement today, we have also seen that there are a
number of areas in which academic research and tool vendors should provide
better techniques, approaches and tools to support cloud developers.

\textbf{Academic Research on Infrastructure Evolution.}
In IaaS clouds, software developers make use of scripting or IaC tools to
formally specify and track infrastructure configuration.
This means that infrastructure evolution is now tracked in version control
systems the same way the evolution of regular code artifacts is. We argue that
this provides \emph{new opportunities for academic research} on software repository
mining to investigate how infrastructure code evolves over the lifetime of a
project, compared to the overall application code.
This will allow us to, for instance, discover anti-patterns in infrastructure provisioning code.

\textbf{Improved Log and Metrics Management.}
Cloud developers need to  anticipate problems prior to deployment
and define relevant logs and operational metrics. While this is largely already possible today, there is little support to guide developers \emph{what} and \emph{how} they should be logging exactly. Some study participants have reported that this results in rather excruciating
trial-and-error. We argue that correct tracing and metric definition has to be introduced
as part of the development workflow. Methods and tooling should be improved to support the process of defining and evolving useful logs and metrics for cloud applications.

\textbf{Local Reproducibility through Containers.}
Cloud applications are distributed by default. Our study participants reported
that trying to \emph{reproduce faults locally} can be a tedious and time-consuming
task. It involves knowing the exact state of when the fault occurred in
production (i.e., infrastructure and data state), as well as the capability to
replicate the environment and its state locally. We envision methods and tools
that have the ability to recreate a distributed environment in a local
development environment from a snapshot of when the fault occurred, utilizing
containerization technology (e.g., using Docker).

\textbf{Tools for Developer Targeted Analytics.}
In the cloud, more metrics are available, both in quantity and dimension.
However, in our study we have observed that developers, before utilizing
existing metrics, much rather rely on their experience and intuition to solve problems.
Besides a required change in mindset, as discussed in Section~\ref{sec:impl}, we
attribute this to the fact that most monitoring tools are built to be used by
operations teams, rather than software developers.
Currently, the existing way of delivering metrics is difficult to leverage for
developers as it is \emph{not actionable in the development process}.
Existing tools need to expose better APIs for data extraction and integration.
Future research needs to address how the abundance of data in the cloud can
become more actionable for developers and integrate it into their daily
workflows. Possible solutions could include integration of data in code views
through IDE plugins or within issue tracking systems.
%\tf{this paragraph related it back nicely, try to do it similarly for most others too if possible.}

%\begin{itemize}
%	\item Ability to more easily recreate the state of distributed environment locally technologies like docker/lxc can enable this. It would also help to address problems of implementing scalability more easily i.e. if we could test that locally

\section{Threats to Validity}
\label{sec:threats}

While we have designed our research as a mixed-mode study to reduce threats to validity as far as possible, there are still a number of limitations inherent in our research design. Primarily, the question arises to what extent the 25 cloud developers we interviewed are representative (\emph{external validity}). However, to mitigate this threat, we have made sure to recruit interview participants that are approximately evenly distributed between smaller companies and larger enterprises, as well as between IaaS and PaaS clouds. There still are a little more interviewees that deploy on PaaS than on IaaS. We think to have mitigated this concern by having more IaaS particpants in the survey, balancing the overall results. Further, our interview participants cover a broad range of experience levels, and work with various kinds of cloud systems on various kinds of applications. 
A similar threat also exists for the external validity of our survey. We recruited our participants almost exclusively via GitHub, meaning that we are likely to have attracted mostly software engineers who are actively interested in open source development, and who are also following the progress of at least one big open source cloud product. Further, responses were necessarily voluntary, hence we are likely to have attracted a crowd with higher-than-average interest in the topic of cloud computing.

In terms of \emph{internal validity}, it is possible that we have biased our interview partners through the pre-selection of questions and topics, and that we missed entirely unanticpiated differences and implications of software development in the cloud. Also, our themes focused on how software development in the cloud is different than in non-cloud environments. Thus, interview participants may have been inclined to overstate differences, leaving out similarities. However, given that no major undiscovered differences and implications were mentioned during the survey either, we judge this threat to be low. Another threat to internal validity is that we had to translate a subset of our interviews to English for analysis, offering the possibility that subtle semantics have changed during translation. We strived to minimize this threat by carefully checking all quotes and translations against the original transcripts, to make sure that the original meaning remained unchanged. %Finally, one can raise the question to what extent our survey participants have truthfully reported on their actual behavior and experiences. The relatively high agreement with most of our observations (e.g., 72\% agreeing that they now look at more data) is suspicious, and may indicate that some participants anticipated what the ``expected'' answer was more than responding truthfully. 
\section{Conclusions}
\label{sec:conc}

We report on the first systematic study on how professional software developers build applications and utilize specialized tools and data on top of cloud infrastructures and platforms. These insights provided by our study help to better understand how cloud computing has made an impact throughout the software development life cycle.
Our findings suggest, among others things, two major developments: (1) developers need better means to anticipate runtime problems in the cloud and rigorously define and collect metrics for better fault localization and (2) the cloud offers an abundance of operational data, however, developers still often rely on their experience and intuition rather than utilizing metrics to solve problems. Methods and tools for developers will, therefore, need to adapt to these required changes in the cloud. From our findings, we extracted a set of guidelines for cloud development and identified challenges for researchers and tool vendors to support developers in these efforts by developing new approaches for managing metrics and making operational data more actionable.

\section*{Acknowledgment}
The authors would like to thank all interview and survey participants. The research leading to these results has received funding from the European Community's Seventh Framework Programme (FP7/2007-2013) under grant agreement no. 610802 (CloudWave).



\bibliographystyle{IEEEtran}
\bibliography{bibtex}

\begin{thebibliography}{10}
\providecommand{\url}[1]{#1}
\csname url@samestyle\endcsname
\providecommand{\newblock}{\relax}
\providecommand{\bibinfo}[2]{#2}
\providecommand{\BIBentrySTDinterwordspacing}{\spaceskip=0pt\relax}
\providecommand{\BIBentryALTinterwordstretchfactor}{4}
\providecommand{\BIBentryALTinterwordspacing}{\spaceskip=\fontdimen2\font plus
\BIBentryALTinterwordstretchfactor\fontdimen3\font minus
  \fontdimen4\font\relax}
\providecommand{\BIBforeignlanguage}[2]{{%
\expandafter\ifx\csname l@#1\endcsname\relax
\typeout{** WARNING: IEEEtran.bst: No hyphenation pattern has been}%
\typeout{** loaded for the language `#1'. Using the pattern for}%
\typeout{** the default language instead.}%
\else
\language=\csname l@#1\endcsname
\fi
#2}}
\providecommand{\BIBdecl}{\relax}
\BIBdecl

\bibitem{buyya:09}
R.~Buyya, C.~S. Yeo, S.~Venugopal, J.~Broberg, and I.~Brandic, ``{Cloud
  Computing and Emerging IT Platforms: Vision, Hype, and Reality for Delivering
  Computing As the 5th Utility},'' \emph{{Future Generation Computer Systems}},
  vol.~25, no.~6, pp. 599--616, 2009.

\bibitem{Armbrust:2010qy}
M.~Armbrust, A.~Fox, R.~Griffith, A.~D. Joseph, R.~Katz, A.~Konwinski, G.~Lee,
  D.~Patterson, A.~Rabkin, I.~Stoica \emph{et~al.}, ``{A View of Cloud
  Computing},'' \emph{{Communications of the ACM}}, vol.~53, no.~4, pp. 50--58,
  2010.

\bibitem{huettermann:12}
M.~H\"uttermann, \emph{{DevOps for Developers}}.\hskip 1em plus 0.5em minus
  0.4em\relax {Apress}, 2012.

\bibitem{beloglazov:12}
A.~Beloglazov, J.~Abawajy, and R.~Buyya, ``{Energy-Aware Resource Allocation
  Heuristics for Efficient Management of Data Centers for Cloud Computing},''
  \emph{{Future Generation Computer Systems}}, vol.~28, no.~5, pp. 755--768,
  May 2012.

\bibitem{marshall:11}
P.~Marshall, K.~Keahey, and T.~Freeman, ``{Improving Utilization of
  Infrastructure Clouds},'' in \emph{{Proceedings of the 2011 11th IEEE/ACM
  International Symposium on Cluster, Cloud and Grid Computing (CCGRID
  '11)}}.\hskip 1em plus 0.5em minus 0.4em\relax Washington, DC, USA: IEEE
  Computer Society, 2011, pp. 205--214.

\bibitem{iosup:11}
A.~Iosup, S.~Ostermann, N.~Yigitbasi, R.~Prodan, T.~Fahringer, and D.~Epema,
  ``{Performance Analysis of Cloud Computing Services for Many-Tasks Scientific
  Computing},'' \emph{{IEEE Transactions on Parallel and Distributed Systems}},
  vol.~22, no.~6, pp. 931--945, Jun. 2011.

\bibitem{barker:14}
A.~Barker, B.~Varghese, J.~S. Ward, and I.~Sommerville, ``{Academic Cloud
  Computing Research: Five Pitfalls and Five Opportunities},'' in
  \emph{{Proceedings of the 6th USENIX Workshop on Hot Topics in Cloud
  Computing (HotCloud 14)}}.\hskip 1em plus 0.5em minus 0.4em\relax
  Philadelphia, PA: USENIX Association, Jun. 2014.

\bibitem{mell:13}
P.~Mell and T.~Grance, ``{The NIST Definition of Cloud Computing},'' National
  Institute of Standards and Technology (NIST), Gaithersburg, MD, Tech. Rep.
  800-145, September 2011.

\bibitem{lawton:08}
G.~Lawton, ``{Developing Software Online With Platform-as-a-Service
  Technology},'' \emph{Computer}, vol.~41, no.~6, pp. 13--15, Jun. 2008.

\bibitem{singer:14}
L.~Singer, F.~Figueira~Filho, and M.-A. Storey, ``{Software Engineering at the
  Speed of Light: How Developers Stay Current Using Twitter},'' in
  \emph{{Proceedings of the 36th International Conference on Software
  Engineering (ICSE 2014)}}.\hskip 1em plus 0.5em minus 0.4em\relax New York,
  NY, USA: ACM, 2014, pp. 211--221.

\bibitem{murphy:13}
E.~Murphy-Hill, T.~Zimmermann, C.~Bird, and N.~Nagappan, ``{The Design of Bug
  Fixes},'' in \emph{{Proceedings of the 2013 International Conference on
  Software Engineering (ICSE'13)}}.\hskip 1em plus 0.5em minus 0.4em\relax
  Piscataway, NJ, USA: IEEE Press, 2013, pp. 332--341.

\bibitem{khajeh:11}
A.~Khajeh-Hosseini, I.~Sommerville, and I.~Sriram, ``{Research Challenges for
  Enterprise Cloud Computing},'' \emph{CoRR}, vol. abs/1001.3257, 2010.

\bibitem{mei:08}
L.~Mei, W.~K. Chan, and T.~H. Tse, ``{A Tale of Clouds: Paradigm Comparisons
  and Some Thoughts on Research Issues},'' in \emph{{Proceedings of the 2008
  IEEE Asia-Pacific Services Computing Conference (APSCC '08)}}.\hskip 1em plus
  0.5em minus 0.4em\relax Washington, DC, USA: IEEE Computer Society, 2008, pp.
  464--469.

\bibitem{mei:13}
L.~Mei, Z.~Zhang, and W.~K. Chan, ``{More Tales of Clouds: Software Engineering
  Research Issues from the Cloud Application Perspective},'' \emph{{Proceedings
  of the 2013 IEEE 37th Annual Computer Software and Applications Conference
  (COMPSAC)}}, vol.~1, pp. 525--530, 2009.

\bibitem{palanisamy:11}
B.~Palanisamy, A.~Singh, L.~Liu, and B.~Jain, ``{Purlieus: Locality-Aware
  Resource Allocation for MapReduce in a Cloud},'' in \emph{{Proceedings of
  2011 International Conference for High Performance Computing, Networking,
  Storage and Analysis (SC'11)}}.\hskip 1em plus 0.5em minus 0.4em\relax New
  York, NY, USA: ACM, 2011, pp. 58:1--58:11.

\bibitem{leitner:12}
P.~Leitner, B.~Satzger, W.~Hummer, C.~Inzinger, and S.~Dustdar, ``{CloudScale:
  A Novel Middleware for Building Transparently Scaling Cloud Applications},''
  in \emph{{Proceedings of the 27th Annual ACM Symposium on Applied Computing
  (SAC'12)}}.\hskip 1em plus 0.5em minus 0.4em\relax New York, NY, USA: ACM,
  2012, pp. 434--440.

\bibitem{jayaram:13}
K.~Jayaram, ``\BIBforeignlanguage{English}{Elastic remote methods},'' in
  \emph{\BIBforeignlanguage{English}{Middleware 2013}}, ser. Lecture Notes in
  Computer Science, D.~Eyers and K.~Schwan, Eds.\hskip 1em plus 0.5em minus
  0.4em\relax Springer Berlin Heidelberg, 2013, vol. 8275, pp. 143--162.

\bibitem{narasimhan:11}
B.~Narasimhan and R.~Nichols, ``{State of Cloud Applications and Platforms: The
  Cloud Adopters' View},'' \emph{Computer}, vol.~44, no.~3, pp. 24--28, 2011.

\bibitem{gupta:13}
P.~Gupta, A.~Seetharaman, and J.~R. Raj, ``{The Usage and Adoption of Cloud
  Computing by Small and Medium Businesses},'' \emph{{International Journal of
  Information Management}}, vol.~33, no.~5, pp. 861 -- 874, 2013.

\bibitem{devops:14}
\BIBentryALTinterwordspacing
``{2014 State of DevOps Report},'' Puppet Labs, IT Revolution Press, and
  ThoughtWorks, Tech. Rep., 2014. [Online]. Available:
  \url{http://puppetlabs.com/2014-devops-report}
\BIBentrySTDinterwordspacing

\bibitem{shiver:14}
R.~Shiver, ``{Survey: Enterprise Development in the Cloud},'' Gigaom Research,
  Tech. Rep., 2014.

\bibitem{hoda:12}
R.~Hoda, J.~Noble, and S.~Marshall, ``{Developing a Grounded Theory to Explain
  the Practices of Self-Organizing Agile Teams},'' \emph{{Empirical Software
  Engineering}}, vol.~17, no.~6, pp. 609--639, 2012.

\bibitem{bratthall:02}
L.~Bratthall and M.~Jorgensen, ``{Can you Trust a Single Data Source
  Exploratory Software Engineering Case Study?}'' \emph{{Empirical Software
  Engineering}}, vol.~7, no.~1, pp. 9--26, 2002.

\bibitem{infneeds}
R.~P. Buse and T.~Zimmermann, ``Information needs for software development
  analytics,'' in \emph{Proceedings of the 34th International Conference on
  Software Engineering}, June 2012.

\bibitem{tang:14}
X.~Tang, Z.~Zhang, M.~Wang, Y.~Wang, Q.~Feng, and J.~Han, ``{Performance
  Evaluation of Light-Weighted Virtualization for PaaS in Clouds},'' in
  \emph{{Algorithms and Architectures for Parallel Processing}}.\hskip 1em plus
  0.5em minus 0.4em\relax Springer, 2014, pp. 415--428.

\bibitem{newmann:15}
P.~Newman, \emph{{Building Microservices}}.\hskip 1em plus 0.5em minus
  0.4em\relax {O'Reilly}, 2015.

\bibitem{cito:14}
J.~Cito, D.~Suljoti, P.~Leitner, and S.~Dustdar, ``{Identifying Root-Causes of
  Web Performance Degradation using Changepoint Analysis},'' in
  \emph{{Proceedings of the 14th International Conference on Web Engineering
  (ICWE)}}.\hskip 1em plus 0.5em minus 0.4em\relax Springer Berlin Heidelberg,
  2014.

\end{thebibliography}

%\begin{thebibliography}{1}


%\end{thebibliography}

\end{document}